\documentclass[default,iicol]{sn-jnl}

\usepackage{amssymb}
\usepackage{bm}
\usepackage{graphicx}
\usepackage{float}
\usepackage{subfigure}
\usepackage{stackengine}
\usepackage{physics}
\usepackage{hyperref}
\usepackage[T1]{fontenc}
\usepackage{amsmath}
\usepackage{amsfonts}
\usepackage{natbib}

\begin{document}

\title{Characterizing the delocalized-localized Anderson phase
  transition based on the system's response to boundary conditions}

\author*[1]{
  \fnm{Mohammad}
  \sur{Pouranvari}
}
\email{m.pouranvari@umz.ac.ir}

\affil*[1]{
  \orgdiv{Department of Solid-State Physics, Faculty of Science},
  \orgname{University of Mazandaran},
  \city{Babolsar},
  \postcode{4741613534},
  \country{Iran}
}

\abstract{A new characterization of the Anderson phase transition,
  based on the response of the system to the boundary conditions is
  introduced. We change the boundary conditions from periodic to
  antiperiodic and look for its effects on the eigenstate of the
  system. To characterize these effects, we use the overlap of the
  states. In particular, we numerically calculate the overlap between
  the ground-state of the system with periodic and antiperiodic
  boundary conditions in one-dimensional models with
  delocalized-localized phase transitions. We observe that the overlap
  is close to one in the localized phase, and it gets appreciably
  smaller in the delocalized phase. In addition, in models with
  mobility edges, we calculate the overlaps between single-particle
  eigenstate with periodic and antiperiodic boundary conditions to
  characterize the entire spectrum. By this single-particle overlap, we can locate the mobility edges
  between delocalized and localized states.  }

\maketitle

\section{Introduction}\label{Introduction}
Characterizing phases and phase transitions is one of the main goals
of condensed matter physics. Introducing a new method to understand
the nature of the phases, also to locate the phase transition point
are always the essential parts of the current research. In general,
the classical interpretations are enough for a description of the
system phases and for the phase transitions.  The story is different
for a zero-temperature phase transition, where quantum fluctuations
become essential and dominate thermal fluctuations. Among quantum
phase transitions, Anderson phase transition\cite{PhysRev.109.1492}
between a delocalized and a localized phase has attracted much
attentions\cite{lagendijk2009fifty, semeghini2015measurement,
  PhysRevB.5.2931}. In this form of phase transition, disorder plays
the central role. For an ideal clean system, which is translationally
invariant, the Bloch waves propagate through the entire system, and
thus system's eigenstate is extended, and the system is in the
metallic phase. Introducing the disorder in the system, which change
the Physics of the system. Interferences of the scattered waves of the
disorders can be destructive and make the system localized. This
localization depends on the dimension of the system. In one- and
two-dimensional systems, any infinitesimal disorder makes the system
localized\cite{RevModPhys.80.1355}. On the other hand, in the
three-dimensional systems, we have a phase transition between
delocalized and localized phases as we increase the disorder
strength. For a small disorder strength, the state of the system is
still delocalized, but when the disorder strength is larger than a
critical value, it will become localized in a small part of the system
\cite{markos2006numerical, doi:10.1080/13642819308215292} (this
critical value depends on the randomness distribution). There are also
one-dimensional models with a \textit{correlated} disorder that
exhibit phase transitions between delocalized and localized
phases\cite{RevModPhys.80.1355, PhysRevLett.82.4062,
  mirlin1996transition}. On the other hand, Hamiltonian's size of
these one-dimensional models correspond to matrices that increases
linearly with the system size.  Since they represent the same nature
of the Anderson localization, compared to those two- and
three-dimensional models with Anderson phase transition, they are
more suitable for numerical calculations.

People use different quantities to characterize the transition between
delocalized and localized phases. Since the eigenstate of the system
at the Fermi level shows the tendency of the material to conduct an
electron, one of the obvious characterizations is to measure the
extent of the Fermi level eigenstate. To quantify how much an
eigenstate of the system $\psi$ is extended, people use the
participation ratio:
\begin{equation}
  \text{PR} = \frac{1}{\sum_{i=1}^N \abs{\psi_i}^4},
\end{equation}
($N$ is the system size) where for a normalized eigenstate, $\psi_i$
is the probability amplitude at each site $i$. In the delocalized
phase, where the system is extended in the entire system, $\psi_i$
approaches $1/\sqrt{N}$, and thus PR goes to $N$. On the other hand,
in the localized phase where the state of the system is localized at a
few sites, PR approaches to $O(1)$. The next candidate would be
entanglement. In the delocalized phase, we expect a larger correlation
in the system than in the localized phase. Thus, the amount of the
entanglement is larger in the delocalized phase. Thus, the system
entanglement properties can locate the phase transition
point\cite{PhysRev.47.777, RevModPhys.73.565, schrodinger_1935,
  Osterloh2002, RevModPhys.80.517, RevModPhys.81.865}.

Besides the system's eigenstate, looking at the system's energy
eigenvalues is also informative. Level spacing (defined as
$\Delta_n = E_{n+1}-E_n$ as a difference between the adjacent energies
$\{E\}$) and their distributions are another way to characterize a
delocalized from a localized phases\cite{PhysRevB.47.11487}. That
stems from the fact that, in contrast to the localized phase, the
energy spectrum is doubly degenerated in the delocalized phase, and
thus there is a gap between even-odd and odd-even level
spacing\cite{aubry1980analyticity, PhysRevLett.123.025301}. Moreover,
the ratio of the level spacing:
\begin{equation}
  r_n = \frac{\text{min}(\Delta_n, \Delta_{n+1})}{\text{max}(\Delta_n, \Delta_{n+1})},
\end{equation}
is also useful. In the delocalized phase, where the energy spectrum
has a Wigner-Dyson distribution, the disorder average of $r_n$
approaches to $\approx 0.53$, and in the localized phase with Poisson
statistics of the energy spectrum, it goes to $\approx 0.386$.  Thus,
people use the ratio of the level spacing to distinguish delocalized
from localized phases\cite{PhysRevB.75.155111, PhysRevLett.110.084101,
  PhysRevB.97.125116}.

The response of a system to a local quench is also another
characterization. A measure of this response is fidelity.  If we
consider $\ket{G}$ as the ground state of the system without a local
quench and $\ket{G'}$ as the ground state of the system with a local
quench, then the fidelity is the overlap $F=\abs{\bra{G}\ket{G'}}$ of
these two states \cite{PhysRevLett.98.110601,PhysRevA.77.032111,
  PhysRevB.76.180403, PhysRevE.98.062137,
  PhysRevB.80.014403,PhysRevA.89.033625}. This overlap goes to zero in
a power-law fashion ($F \sim N^{-\gamma}$) in the delocalized phase,
the so-called Anderson orthogonality catastrophe
\cite{PhysRevLett.18.1049}. While, it decays exponentially in the
localized phase ($F\sim e^{-\beta N}$); where $\gamma$ and $\beta$
depend on the disorder strength\cite{PhysRevLett.18.1049,
  PhysRevB.92.054203, PhysRevB.92.220201,Cosco_2018,
  PhysRevLett.122.040604}.

To characterize the delocalized-localized phase transition, we can
also look at the system's behavior upon the change in the boundary
conditions\cite{PhysRevLett.93.266402, Peschel_2005}. For an extended
eigenstate, the change in the boundary conditions is seen by the
eigenstate, so it is reflected in the corresponding eigenenergy. In
contrast, in a localized eigenstate, where the amplitude of the
wavefunction is approximately non-zero for some finite number of sites
only, the change in the boundary conditions is not seen by the
eigenstate. Thus, there will be no change in the corresponding
eigenenergy. On this subject, Ref. \cite{Edwards_1972} used the shift
in the eigenenergy when the boundary conditions are changed from
periodic to antiperiodic to characterize the delocalized-localized
phase transition. In the same way, in Ref \cite{PhysRevB.96.045123},
we change the boundary conditions from periodic to antiperiodic and
calculate the shift in the entanglement spectrum, and also the shift
in the entanglement entropy. We observe that the entanglement
properties of the system are sensitive to the boundary conditions in
the delocalized phase, and become insensitive in the localized
phase. Thus, the shift in the entanglement entropy and spectrum can be
used to characterize the Anderson phase transition. Similarly, we
studied the change in the single-particle density matrix for a
many-body system, when boundary conditions are changed from periodic
to antiperiodic in Ref \cite{PhysRevB.103.035136}. We observed that
the shift in the spectrum of the single-particle density matrix is
non-vanishing in the delocalized phase, and it goes to zero in the
localized phase; it is thus a characterization of the many-body
localization.

In this paper, we look for the \textit{direct} effect of the change in
the boundary conditions on the system's \textit{state} (We know that
we can directly observe the eigenstate of the system in
experiment\cite{PhysRevLett.101.256802}). In practice, we use the
concept of the fidelity regarding the states with different boundary
conditions. In particular, we calculate the overlap between the state
of the system with the periodic boundary conditions (PBC) and the
corresponding state of the system with antiperiodic boundary
conditions (APBC) to characterize their similarity. We expect that the
overlap (between the state of the system with PBC and APBC) becomes
unity in the localized phase and it becomes smaller than one in the
delocalized phase. In this regard, we expect that the overlap
distinguishes a localized from a delocalized phases.

In free fermion models, to obtain the ground state of the system
corresponding to a Fermi energy, we do the followings. We fill
single-particle eigenstates of the system from the lowest eigen-energy
up to the Fermi level, and the many-body ground state of the system is
the Slater determinant of these single-particle eigenstates. To
observe and calculate the similarity between many-body eigenstates, we
can thus calculate the overlap between the many-body ground state of
the system with PBC and APBC. Besides, we can also look at the
overlaps of the single-particle eigenstates separately---those
eigenstates that build the many-body state of the system. In this
regard, we calculate the many-body ground states of the system with
PBC and APBC, then we obtain their overlap, and we call it
ground-state overlap (GSO). In addition, we calculate the
single-particle eigenstates of the system with PBC and APBC, and we
call their corresponding overlaps single-particle overlap (SPO).  We
show that the behavior of the GSO and SPO are different in the
delocalized and the localized phases. Thus, we utilize them to
characterize the delocalized-localized phase transition.

The remainder of this paper is as follows. In section \ref{method} we
explain the models we employ in this paper to verify our ideas. Also
the calculations methods for the SPO and GSO are explained. Section
\ref{result} is devoted to our numerical calculations, where we
present the results for the SPO and GSO for the models. We also show
how they can be used as phase transition characterization. We conclude
the paper in section \ref{conclusion} with some suggestions for future
works.

\section{Method and models}\label{method}
We work with one-dimensional free fermion tight-binding models with
the following Hamiltonian:
\begin{equation}\label{hamiltonian}
  H=-t\sum_{i=1}^{N} (c_i^{\dagger} c_{i+1}+c_{i+1}^{\dagger} c_i)+
  \sum_{i=1}^{N} \varepsilon_i c_i^{\dagger} c_i,
\end{equation}
where, $c_i(c_i^{\dagger})$ is the annihilation (creation) operator
for the $i$th site. $N$ is the number of sites in the system. The
amplitude of the nearest-neighbor hopping is $t$, and we set it to be
$1$ as the energy scale. The models we use in this paper are
determined by their on-site energies $\{\varepsilon\}$. One of the
models is the random dimer (RD) model, where $\varepsilon_i$ are
chosen randomly to be either of the two choices of $\phi_a$ and
$\phi_b$. One of them (here $\phi_b$) is attributed to two successive
sites (so this model is called dimer). It is
shown\cite{PhysRevLett.65.88} that states at the resonant energy
$E=\phi_b$ are delocalized when $-2t \le \phi_a-\phi_b \le 2t$. In
this study, we choose $\phi_a=0$, so the condition is
$-2 \le \phi_b \le 2$. States at energies other than the resonant
energy are localized. The Anderson phase transition in this model has
been studied before from different perspectives\cite{Bovier_1992,
  PhysRevB.69.085109, PhysRevB.48.16347, PhysRevB.56.1170,
  PhysRevB.100.195109, PhysRevB.96.045123}.

Another model is the generalized Aubry-Andry
model\cite{PhysRevLett.114.146601}, where the on-site energies
$\varepsilon_i$ are:

\begin{equation}\label {gAA}
  \varepsilon_i = 2 \lambda \frac{\cos{(2\pi i b)}}{1-\alpha\cos{(2\pi i b)}},
\end{equation}
$b$ is an irrational number, and we set it to be the golden ratio
$b = \frac{1+\sqrt{5}}{2}$. Since $b$ is not a rational number, the
system has incommensurate periodicity with respect to the lattice
periodicity, which we set it to be $1$. Thus, this system is neither
completely periodic nor completely random. This model has mobility
edges separating delocalized and localized eigenstates at the
following eigenenergy\cite{PhysRevLett.114.146601}:
\begin{equation}\label{me}
  E_{\text{mobility edge}} = 2 sgn(\lambda)\frac{\abs{t}-\abs{\lambda}}{\alpha}.
\end{equation}

One special case for this model is $\alpha=0$, which is the
Aubry-Andry model, with the following on-site energies:

\begin{equation}\label{varepsilonAA}
  \varepsilon_i = 2 \lambda \cos(2\pi i b + \theta),
\end{equation}
here, a random phase $\theta$ is added which is distributed uniformly
between $-\pi$ and $\pi$.  It is shown that for $\lambda<1$, all
states are delocalized and for $\lambda>1$, all states are
localized\cite{aubry1980analyticity}. Thus, there is a phase
transition between delocalized and localized phases at
$\lambda=1$. The properties of this model have been studied
before\cite{Dom_nguez_Castro_2019, PhysRevB.101.174203,
  PhysRevB.101.024202,PhysRevB.100.195109}. We should emphasize that
all the models mentioned above, and their Anderson phase transition
have been studied thoroughly before; here we just use them to verify
our idea.

Since both models describe the free fermions, we deal with Hamiltonians
that are represented by $N \times N$ matrices. These matrices can be
diagonalized numerically. In this paper, we use LAPACK\cite{laug} to
diagonalize the matrices, and obtain their eigenvalues and
eigenvectors.

In this paper, we want to consider the overlap of the state of the
system with PBC and the state of the system with APBC. Here, we
explain how to calculate the SPO and GSO overlaps. By PBC we mean
$c_{N+1}^{\dagger} =+ c_1^{\dagger}$, and by APBC we mean
$c_{N+1}^{\dagger} = -c_1^{\dagger}$. If we assume the following free
fermion Hamiltonians with PBC and APBC:

\begin{eqnarray}
  H_{\text{PBC}} &=& \sum_{i,j}^N h_{ij}^{\text{P}} c_i^{\dagger} c_j,\\
  H_{\text{APBC}} &=& \sum_{i,j}^N h_{ij}^{\text{A}} c_i^{\dagger} c_j,
\end{eqnarray}
where, $h^{\text{P}}$ and $h^{\text{A}}$ can be determined by the
choice of the on-site energies (RD, gAA, or AA model) and the boundary
conditions (either PBC or APBC). We can diagonalize the matrix $h$ in the
following way:
\begin{eqnarray}
  h^{\text{P}} &=& UE^{\text{P}}U^{\dagger}\\
  h^{\text{A}} &=& VE^{\text{A}}V^{\dagger},
\end{eqnarray}
and find the eigen-energies $E$ as well as the single-particle
eigenstates $\psi$ for each Hamiltonian
($\psi_i^{n, \text{PBC}} = U_{in}$ is the $i$th element of the $n$th
eigenvector, and similarly for $\psi_i^{n, \text{APBC}} = V_{in}$). To
calculate the ground-state overlap, which is the overlap between
ground-state of the Hamiltonian with the PBC
($\psi^{\text{PBC}}_{MB}$) and APBC ($\psi^{\text{PBC}}_{MB}$), we use
the following method\cite{PhysRevB.92.220201}. After diagonalization,
we can write both Hamiltonians as:
\begin{eqnarray}
  H_{\text{PBC}} &=& \sum_{k=1}^N E_k^{\text{P}} b_k^{\dagger} b_k,\\
  H_{\text{APBC}} &=& \sum_{k=1}^N E_k^{\text{A}} a_k^{\dagger} a_k,
\end{eqnarray}
where $b^{\dagger}_k=\sum_i U_{ik}c^{\dagger}_i$, and
$a^{\dagger}_k=\sum_i V_{ik}c^{\dagger}_i$. Then, GSO is:
\begin{align}
  \text{GSO} &=&\abs{ \bra{\psi^{\text{PBC}}_{MB}}\ket{\psi^{\text{APBC}}_{MB}} } \nonumber \\
             &=& \abs{\bra{0} \prod_{k}^{N_F} b_k \prod_{k'}^{N_F} a_k^{\dagger} \ket{0}}\nonumber \\
             &=& \abs{\det(B)} \label{B}
\end{align}
where $B$ is a matrix built from the first $N_F \times N_F$
part of the matrix $U^{\dagger}V$ ($N_F$ is the number of fermions).

We also use the notion of the single-particle overlap, which is the
overlap between corresponding single-particle eigenstates of the
Hamiltonian with PBC and with APBC. To calculate it, we dot product
the single-particle eigenstate $\psi_{n}^{\text{APBC}}$ with the
corresponding eigenstate $\psi_{n}^{\text{PBC}}$ for specific $n$th
level. We consider only its absolute values:
\begin{equation}
  \text{SPO} =\abs{ \bra{\psi_{n}^{\text{PBC}}}\ket{\psi_{n}^{\text{APBC}}}}
\end{equation}
We note that there is randomness in the AA and RD models, and thus we
take the disorder average of the above-mentioned GSO and SPO over
different random realizations to obtain their mean values.

\section{Results}\label{result}
In this section, we present the results of the numerical
calculations. First, we will study the GSO for the above mentioned
models. In the subsequent subsection, we present the results of the
SPO and we explain its benefits over GSO for models with mobility
edges.

\subsection{Ground-State Overlap}
\begin{figure}
  \centering
  \includegraphics[width=0.48\textwidth]{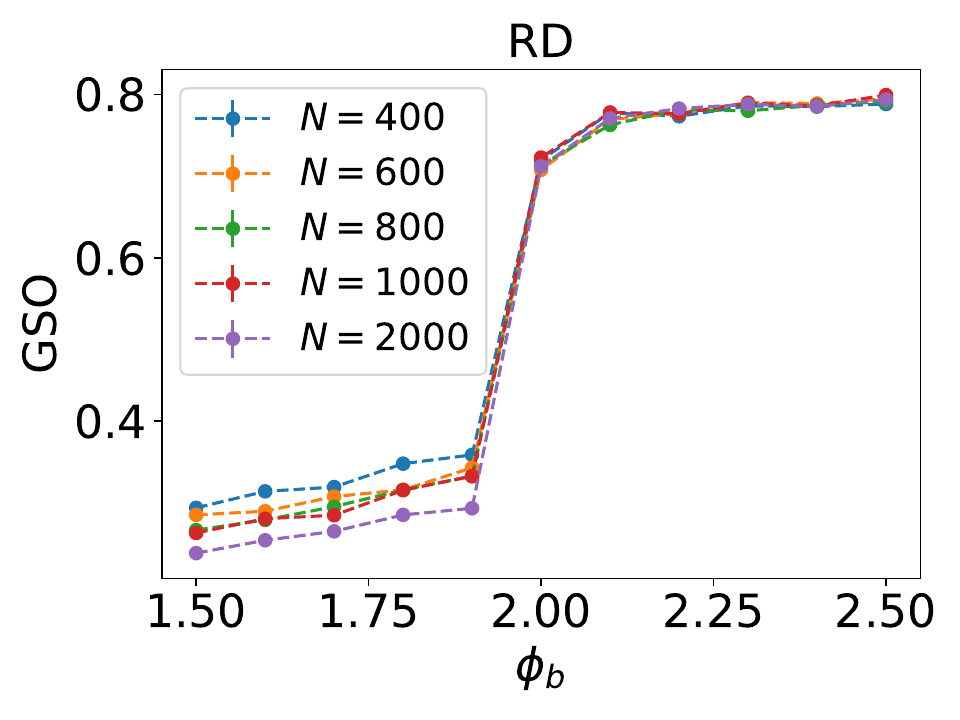}
  \caption{The overlap between the ground-state of the system with PBC
    and APBC, as we change $\phi_b$ for the RD model.  We set
    $E_F=\phi_b$. At each data point, the disorder average is taken
    over $2000$ random realizations. Behavior of the GSO for this
    model is sharp at the phase transition
    point. \label{fig:RD_absoverlap_MB_vsphib}}
\end{figure}

\begin{figure}
  \centering
  \includegraphics[width=0.48\textwidth]{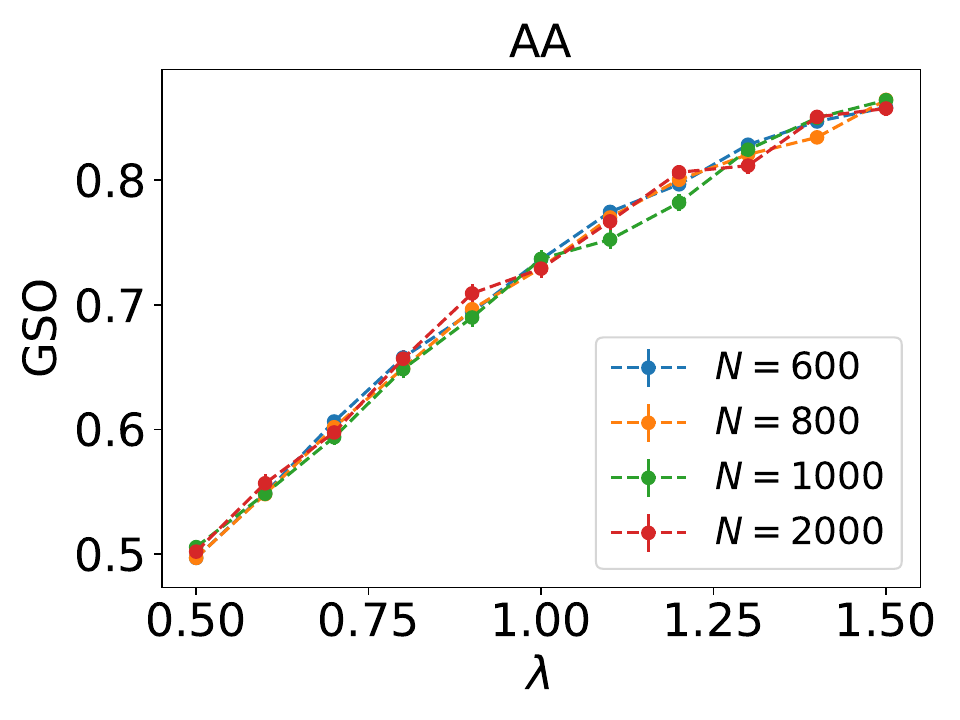}
  \caption{The overlap between the ground-state of the system with PBC
    and APBC, as we change $\lambda$ for the AA model. The GSO is not
    as sharp as in the RD models. The overlaps approaches 1 in the
    localized phase ($\lambda>1$), and it gets smaller in the
    delocalized phase ($\lambda<1$). We set $E_F=0$. At each data
    point, the disorder average is taken over $2000$ random
    realizations. \label{fig:AA_absoverlap_MB_vsLAMBDA} }
\end{figure}

First we calculate the GSO for the RD model, we as we change the
$\phi_b$ and we always set $E_F=\phi_b$. The state of the system at
$E=E_F$ is delocalized for $\phi_b<2$, and it is localized for
$\phi_b>2$ (because of the mirror symmetry in this model, we only
consider the positive part). The results are plotted in the
Fig. \ref{fig:RD_absoverlap_MB_vsphib} for different system sizes
$N$. We can see that GSO has different behaviors in the delocalized
and localized phases. In the delocalized phase ($\phi_b<2$), the GSO
is smaller compared with that in the localized phase ($\phi_b<2$). We
can also see that it approaches unity deep in the localized phase. It
is evident that we can distinguish delocalized and localized phases
from the behavior of the GSO near the phase transition point. We also
calculate the GSO for the AA model. The results are plotted in
Fig. \ref{fig:AA_absoverlap_MB_vsLAMBDA}. The behavior of the GSO for
the AA model is not as sharp as the behavior of the RD model. But,
still it is obvious that GSO approaches $1$ in the localized phase
($\lambda>1$), and it is smaller in the delocalized phase,
($\lambda<1$).

\subsection{Single-Particle Overlap}
\begin{figure}
  \centering
  \begin{subfigure}{}%
    \includegraphics[width=0.23\textwidth]{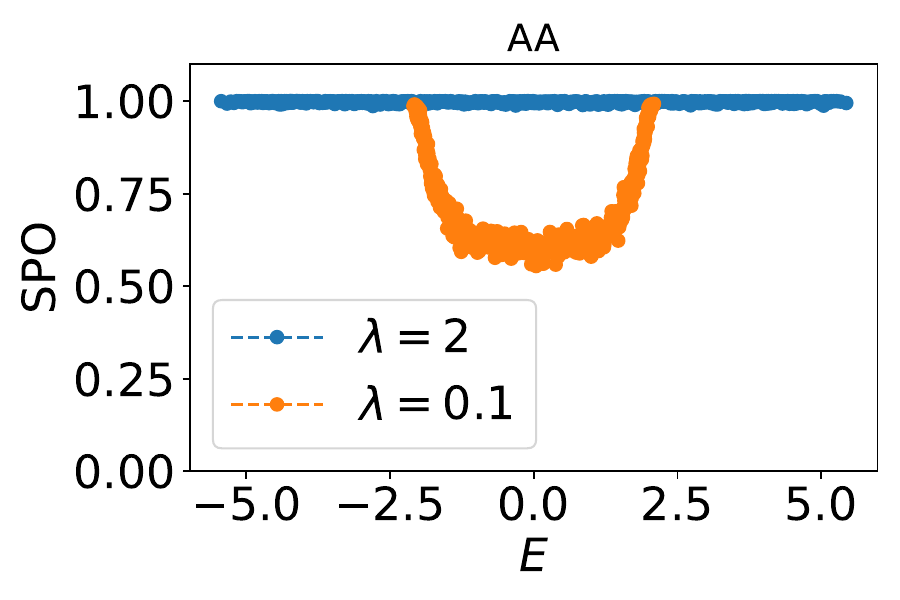}
  \end{subfigure}% ~%
  \begin{subfigure}{}%
    \includegraphics[width=0.23\textwidth]{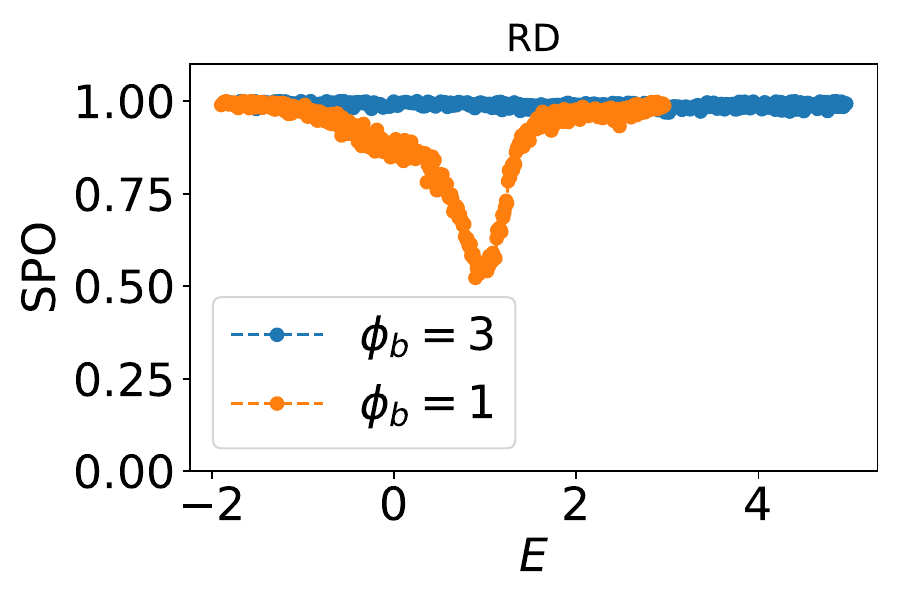}
  \end{subfigure}% ~%
  \caption{The single-particle overlap between the single-particle
    eigenstates with PBC and APBC for the entire spectrum. Left panel:
    SPO for the AA model. In the localized phase ($\lambda=2$) SPO is
    very close to $1$ for the entire spectrum; in the delocalized
    phase ($\lambda=0.1$) SPO is lower than $1$ for the entire
    spectrum. Right panel: SPO for the RD model. For $\phi_b=3$ where
    all single-particle eigenstates are localized, SPO is close to $1$
    for the entire spectrum. For the $\phi_b=1$ with a delocalized
    single-particle eigenstate at the $E=1$, SPO for points close to
    this energy is lower than $1$. \label{fig:AA_RD} }
\end{figure}

To calculate the GSO for the AA and RD models, first we set a Fermi
energy, and then based on the obtained number of fermions, we use
Eq. (\ref{B}). Based on the GSO, we saw that there is a distinction
between delocalized and localized states for both models. let's look
at the the entire spectrum in these models. We know that the AA model
does not have mobility edges between delocalized and localized
states. Either all the states are delocalized ($\lambda <1$) or all
the states are localized ($\lambda>1$). In the RD model, only the
single-particle eigenstate at the resonant energy is delocalized, and
all the other states are localized. To distinguish between delocalized
and localized \emph{single-particle eigenstates}, we use the notion of
the SPO. For both AA and RD models, we calculated the SPO. The
numerical results are plotted in Fig. \ref{fig:AA_RD}. For the AA
model with either \emph{all} delocalized or \emph{all} localized
states, we see that SPO is very close to $1$ (in the localized phase)
or lower than $1$ (in the delocalized phase). SPO for the RD model has
more features. For the case of $\phi_b=3$, where all states are
localized, we see that SPO is $1$ for the entire spectrum. On the
other hand, for the $\phi_b=1$ where single-particle eigenstate at the
resonant energy $E=\phi_b=1$ is delocalized and all the other
single-particle eigenstates are localized, we see that SPO is smaller
than $1$ around the resonant energy, and it becomes close to $1$ away
from the resonant energy. Thus, the SPO can be used to characterize
the \emph{spectral resolution} of the system for delocalized-localized
phase transition.

\begin{figure}
  \centering
  \begin{subfigure}{}%
    \includegraphics[width=0.23\textwidth]{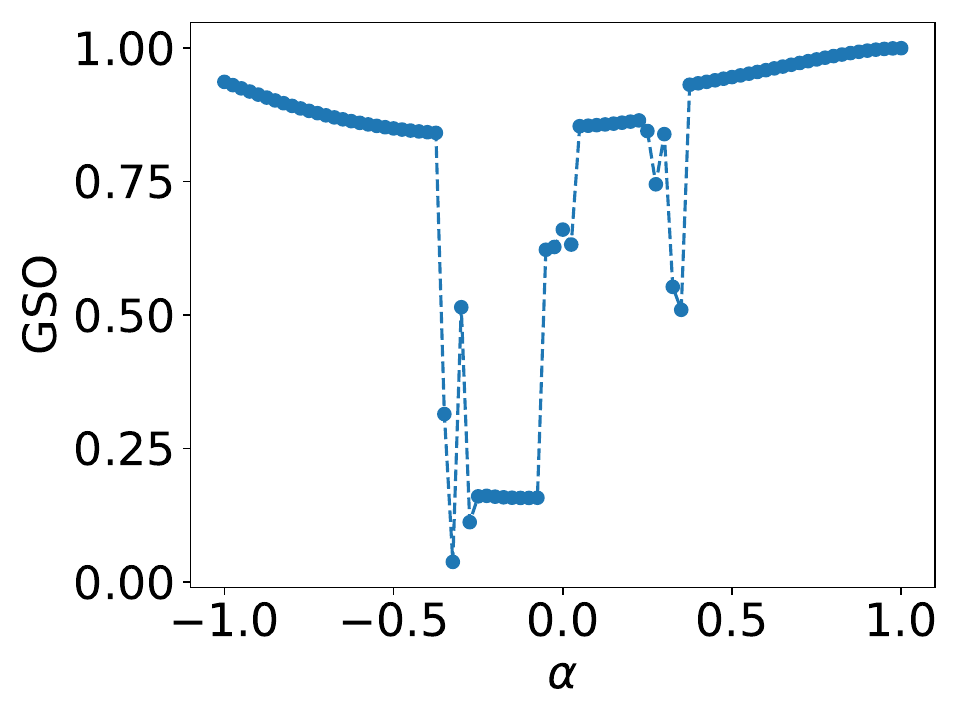}
  \end{subfigure}% ~%
  \begin{subfigure}{}%
    \includegraphics[width=0.23\textwidth]{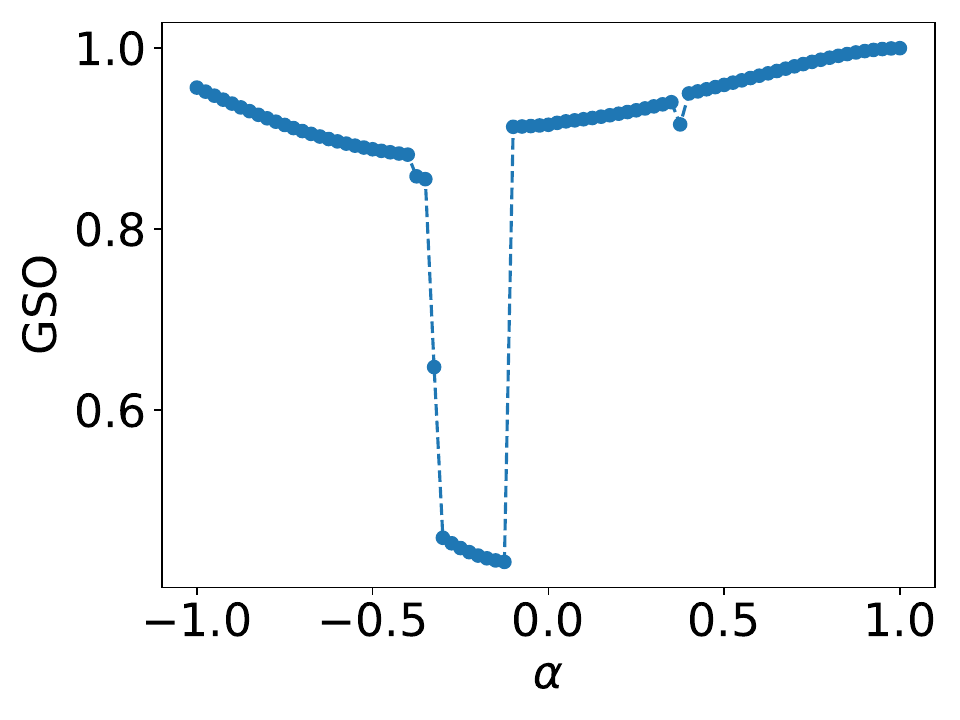}
  \end{subfigure}% ~%
  \caption{the ground-state overlap for the gAA model (Eq. (\ref{gAA})) as
    we change $\alpha$ from $-1$ to $1$ and for two choices of
    $\lambda=0.9$ (left panel) and $\lambda=-1.1$ (right panel). We set
    $N=500, E_F=0$. Although this model has mobility edges, no information about them can be obtained from GSO plots. To see the mobility edges, we use SPO (see Fig. \ref{fig:AAH})
    \label{fig:AA_RD_GSO}}
\end{figure}

It gets more complicated if we consider models with mobility edges
between delocalized and localized phases. One example would be the gAA
model with on-site energies given by Eq. (\ref{gAA}). This model has
mobility edges (given by Eq. (\ref{me})) that separate delocalized
and localized single-particle eigenstates. If we set the Fermi energy
$E_F=0$ and obtain the GSO, we will obtain the plots in
Fig. \ref{fig:AA_RD_GSO}. It is evident that we can not locate the
mobility edges from this plot.

\begin{figure*}
  \centering
  \begin{subfigure}{}%
    \includegraphics[width=0.45\textwidth]{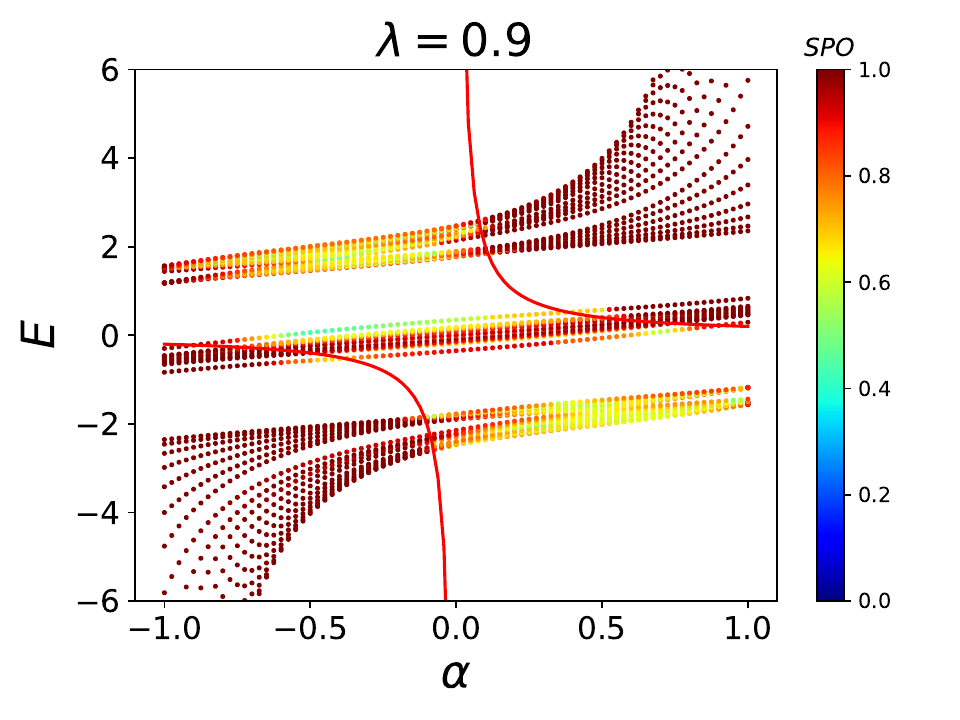}
  \end{subfigure}% ~%
  \begin{subfigure}{}%
    \includegraphics[width=0.45\textwidth]{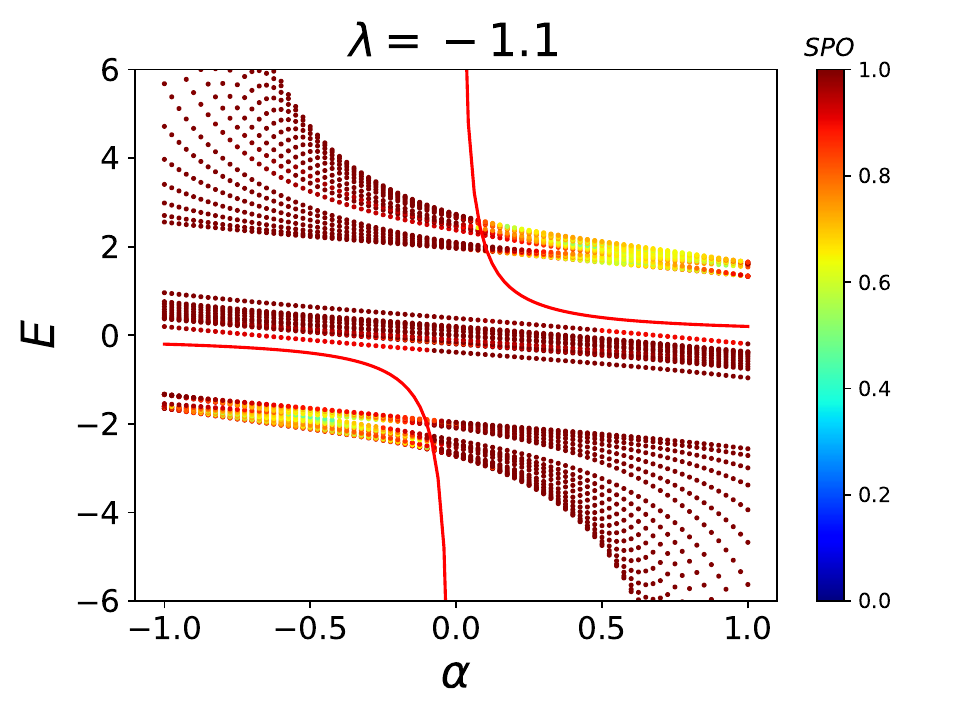}
  \end{subfigure}% ~%
  \caption{ The single-particle overlap for the gAA model
    (Eq. (\ref{gAA})) for the entire spectrum of the Hamiltonian, as
    we change $\alpha$ from $-1$ to $1$ and for two choices of
    $\lambda=0.9$ (left panel) and $\lambda=-1.1$ (right panel). The
    red line is the mobility edges, separating delocalized and
    localized phases based on Eq. (\ref{me}). Color bar shows the
    scale of the SPO.  We set $N=500$. In contrast to GSO
    (Fig. \ref{fig:AA_RD_GSO}), we can locate the mobility edges by
    SPO.
    \label{fig:AAH}}
\end{figure*}

However, the SPO exhibit the detailed features of the mobility
edges. In Fig. \ref{fig:AAH}, we plot the result of the SPO for the
entire spectrum, for $-1 \le \alpha \le 1$ and for two choices of
$\lambda=0.9, -1.1$. Categorization of the single-particle eigenstates
based on the SPO is entirely in agreement with the mobility edges
given by Eq. (\ref{me}). This shows that the SPO between
single-particle eigenstates with PBC and APBC is an informative
measure to characterize the single-particle delocalized and localized
eigenstates.

\section{conclusion}\label{conclusion}
In this paper, we studied the effect of the boundary conditions on the
overlap between the ground-state of the system with PBC and APBC. We
observe that the overlap in the delocalized phase is smaller than the
overlap in the localized phase, where it goes to unity.  These
observations stem from the fact that state of the system in the
localized phase, does not change upon changing the boundary
conditions. However, the single-particle eigenstate is affected by the
change in the boundary conditions in the delocalized phase.

This conjecture enabled us to use the notion of the GSO to distinguish
delocalized from localized phases. For the AA and RD models, we saw
that GSO has distinguished features in the delocalized and localized
phases. In addition, we utilize the notion of the SPO, to characterize
the single-particle mobility edges in models like gAA. This idea can
also be used for characterizing other free fermion models with
mobility edges that separate delocalized, localized, and also
multi-fractal eigenstates\cite{fraxanet2022localization, PhysRevLett.123.025301, PhysRevB.83.075105}. It is also possible to
generalize the notion of the SPO and GSO for many-body interacting
models, although some effort has been done
before\cite{PhysRevB.103.035136}.

\subsection*{Data Availability Statement}
The data-sets generated during and/or analyzed during the current
study are available from the corresponding author on reasonable
request.

\subsection*{Author Contribution Statement }
The conceptualization, numerical calculations, and writing made by the
single author.
\section*{acknowledgments}
The author gratefully acknowledges the high performance
computing center of the university of Mazandaran for providing computing
resources and time.

\bibliographystyle{sn-basic}

\begin{thebibliography}{10}
  \expandafter\ifx\csname url\endcsname\relax
    \def\url#1{\burl{#1}}\fi
  \expandafter\ifx\csname urlprefix\endcsname\relax\def\urlprefix{URL }\fi
  \providecommand{\bibinfo}[2]{#2}
  \providecommand{\eprint}[2][]{\url{#2}}
  \providecommand{\doi}[1]{\url{https://doi.org/#1}}
  \bibcommenthead

\bibitem{PhysRev.109.1492}
  \bibinfo{author}{Anderson, P.~W.}
  \newblock \bibinfo{title}{Absence of diffusion in certain random lattices}.
  \newblock \emph{\bibinfo{journal}{Phys. Rev.}} \textbf{\bibinfo{volume}{109}},
  \bibinfo{pages}{1492--1505} (\bibinfo{year}{1958}).
  \newblock \urlprefix\url{https://link.aps.org/doi/10.1103/PhysRev.109.1492}.
  \newblock \doi{10.1103/PhysRev.109.1492} .

\bibitem{lagendijk2009fifty}
  \bibinfo{author}{Lagendijk, A.}, \bibinfo{author}{Van~Tiggelen, B.} \&
  \bibinfo{author}{Wiersma, D.~S.}
  \newblock \bibinfo{title}{Fifty years of anderson localization}.
  \newblock \emph{\bibinfo{journal}{Phys. Today}}
  \textbf{\bibinfo{volume}{62}}~(8), \bibinfo{pages}{24--29}
  (\bibinfo{year}{2009}) .

\bibitem{semeghini2015measurement}
  \bibinfo{author}{Semeghini, G.} \emph{et~al.}
  \newblock \bibinfo{title}{Measurement of the mobility edge for 3d anderson
    localization}.
  \newblock \emph{\bibinfo{journal}{Nature Physics}}
  \textbf{\bibinfo{volume}{11}}~(7), \bibinfo{pages}{554--559}
  (\bibinfo{year}{2015}) .

\bibitem{PhysRevB.5.2931}
  \bibinfo{author}{Economou, E.~N.} \& \bibinfo{author}{Cohen, M.~H.}
  \newblock \bibinfo{title}{Existence of mobility edges in anderson's model for
    random lattices}.
  \newblock \emph{\bibinfo{journal}{Phys. Rev. B}} \textbf{\bibinfo{volume}{5}},
  \bibinfo{pages}{2931--2948} (\bibinfo{year}{1972}).
  \newblock \urlprefix\url{https://link.aps.org/doi/10.1103/PhysRevB.5.2931}.
  \newblock \doi{10.1103/PhysRevB.5.2931} .

\bibitem{RevModPhys.80.1355}
  \bibinfo{author}{Evers, F.} \& \bibinfo{author}{Mirlin, A.~D.}
  \newblock \bibinfo{title}{Anderson transitions}.
  \newblock \emph{\bibinfo{journal}{Rev. Mod. Phys.}}
  \textbf{\bibinfo{volume}{80}}, \bibinfo{pages}{1355--1417}
  (\bibinfo{year}{2008}).
  \newblock \urlprefix\url{https://link.aps.org/doi/10.1103/RevModPhys.80.1355}.
  \newblock \doi{10.1103/RevModPhys.80.1355} .

\bibitem{markos2006numerical}
  \bibinfo{author}{Markos, P.}
  \newblock \bibinfo{title}{Numerical analysis of the anderson localization}.
  \newblock \emph{\bibinfo{journal}{arXiv preprint cond-mat/0609580}}
  (\bibinfo{year}{2006}) .

\bibitem{doi:10.1080/13642819308215292}
  \bibinfo{author}{Markos, P.} \& \bibinfo{author}{Kramer, B.}
  \newblock \bibinfo{title}{Statistical properties of the anderson transition
    numerical results}.
  \newblock \emph{\bibinfo{journal}{Philosophical Magazine B}}
  \textbf{\bibinfo{volume}{68}}~(3), \bibinfo{pages}{357--379}
  (\bibinfo{year}{1993}).
  \newblock \urlprefix\url{https://doi.org/10.1080/13642819308215292}.
  \newblock \doi{10.1080/13642819308215292},
  \bibinfo{eprint}{{\href{https://arxiv.org/abs/https://doi.org/10.1080/13642819308215292}{{https://doi.org/10.1080/13642819308215292}}}}
  .

\bibitem{PhysRevLett.82.4062}
  \bibinfo{author}{Izrailev, F.~M.} \& \bibinfo{author}{Krokhin, A.~A.}
  \newblock \bibinfo{title}{Localization and the mobility edge in one-dimensional
    potentials with correlated disorder}.
  \newblock \emph{\bibinfo{journal}{Phys. Rev. Lett.}}
  \textbf{\bibinfo{volume}{82}}, \bibinfo{pages}{4062--4065}
  (\bibinfo{year}{1999}).
  \newblock \urlprefix\url{https://link.aps.org/doi/10.1103/PhysRevLett.82.4062}.
  \newblock \doi{10.1103/PhysRevLett.82.4062} .

\bibitem{mirlin1996transition}
  \bibinfo{author}{Mirlin, A.~D.}, \bibinfo{author}{Fyodorov, Y.~V.},
  \bibinfo{author}{Dittes, F.-M.}, \bibinfo{author}{Quezada, J.} \&
  \bibinfo{author}{Seligman, T.~H.}
  \newblock \bibinfo{title}{Transition from localized to extended eigenstates in
    the ensemble of power-law random banded matrices}.
  \newblock \emph{\bibinfo{journal}{Physical Review E}}
  \textbf{\bibinfo{volume}{54}}~(4), \bibinfo{pages}{3221}
  (\bibinfo{year}{1996}) .

\bibitem{PhysRev.47.777}
  \bibinfo{author}{Einstein, A.}, \bibinfo{author}{Podolsky, B.} \&
  \bibinfo{author}{Rosen, N.}
  \newblock \bibinfo{title}{Can quantum-mechanical description of physical
    reality be considered complete?}
  \newblock \emph{\bibinfo{journal}{Phys. Rev.}} \textbf{\bibinfo{volume}{47}},
  \bibinfo{pages}{777--780} (\bibinfo{year}{1935}).
  \newblock \urlprefix\url{https://link.aps.org/doi/10.1103/PhysRev.47.777}.
  \newblock \doi{10.1103/PhysRev.47.777} .

\bibitem{RevModPhys.73.565}
  \bibinfo{author}{Raimond, J.~M.}, \bibinfo{author}{Brune, M.} \&
  \bibinfo{author}{Haroche, S.}
  \newblock \bibinfo{title}{Manipulating quantum entanglement with atoms and
    photons in a cavity}.
  \newblock \emph{\bibinfo{journal}{Rev. Mod. Phys.}}
  \textbf{\bibinfo{volume}{73}}, \bibinfo{pages}{565--582}
  (\bibinfo{year}{2001}).
  \newblock \urlprefix\url{https://link.aps.org/doi/10.1103/RevModPhys.73.565}.
  \newblock \doi{10.1103/RevModPhys.73.565} .

\bibitem{schrodinger_1935}
  \bibinfo{author}{Schr\"{o}dinger, E.}
  \newblock \bibinfo{title}{Discussion of probability relations between separated
    systems}.
  \newblock \emph{\bibinfo{journal}{Mathematical Proceedings of the Cambridge
      Philosophical Society}} \textbf{\bibinfo{volume}{31}}~(4),
  \bibinfo{pages}{555–563} (\bibinfo{year}{1935}).
  \newblock \doi{10.1017/S0305004100013554} .

\bibitem{Osterloh2002}
  \bibinfo{author}{Osterloh, A.}, \bibinfo{author}{Amico, L.},
  \bibinfo{author}{Falci, G.} \& \bibinfo{author}{Fazio, R.}
  \newblock \bibinfo{title}{Scaling of entanglement close to a quantum phase
    transition}.
  \newblock \emph{\bibinfo{journal}{Nature}}
  \textbf{\bibinfo{volume}{416}}~(6881), \bibinfo{pages}{608--610}
  (\bibinfo{year}{2002}).
  \newblock \urlprefix\url{https://doi.org/10.1038/416608a}.
  \newblock \doi{10.1038/416608a} .

\bibitem{RevModPhys.80.517}
  \bibinfo{author}{Amico, L.}, \bibinfo{author}{Fazio, R.},
  \bibinfo{author}{Osterloh, A.} \& \bibinfo{author}{Vedral, V.}
  \newblock \bibinfo{title}{Entanglement in many-body systems}.
  \newblock \emph{\bibinfo{journal}{Rev. Mod. Phys.}}
  \textbf{\bibinfo{volume}{80}}, \bibinfo{pages}{517--576}
  (\bibinfo{year}{2008}).
  \newblock \urlprefix\url{https://link.aps.org/doi/10.1103/RevModPhys.80.517}.
  \newblock \doi{10.1103/RevModPhys.80.517} .

\bibitem{RevModPhys.81.865}
  \bibinfo{author}{Horodecki, R.}, \bibinfo{author}{Horodecki, P.},
  \bibinfo{author}{Horodecki, M.} \& \bibinfo{author}{Horodecki, K.}
  \newblock \bibinfo{title}{Quantum entanglement}.
  \newblock \emph{\bibinfo{journal}{Rev. Mod. Phys.}}
  \textbf{\bibinfo{volume}{81}}, \bibinfo{pages}{865--942}
  (\bibinfo{year}{2009}).
  \newblock \urlprefix\url{https://link.aps.org/doi/10.1103/RevModPhys.81.865}.
  \newblock \doi{10.1103/RevModPhys.81.865} .

\bibitem{PhysRevB.47.11487}
  \bibinfo{author}{Shklovskii, B.~I.}, \bibinfo{author}{Shapiro, B.},
  \bibinfo{author}{Sears, B.~R.}, \bibinfo{author}{Lambrianides, P.} \&
  \bibinfo{author}{Shore, H.~B.}
  \newblock \bibinfo{title}{Statistics of spectra of disordered systems near the
    metal-insulator transition}.
  \newblock \emph{\bibinfo{journal}{Phys. Rev. B}} \textbf{\bibinfo{volume}{47}},
  \bibinfo{pages}{11487--11490} (\bibinfo{year}{1993}).
  \newblock \urlprefix\url{https://link.aps.org/doi/10.1103/PhysRevB.47.11487}.
  \newblock \doi{10.1103/PhysRevB.47.11487} .

\bibitem{aubry1980analyticity}
  \bibinfo{author}{Aubry, S.} \& \bibinfo{author}{Andr{\'e}, G.}
  \newblock \bibinfo{title}{Analyticity breaking and anderson localization in
    incommensurate lattices}.
  \newblock \emph{\bibinfo{journal}{Ann. Israel Phys. Soc}}
  \textbf{\bibinfo{volume}{3}}~(133), \bibinfo{pages}{18}
  (\bibinfo{year}{1980}) .

\bibitem{PhysRevLett.123.025301}
  \bibinfo{author}{Deng, X.}, \bibinfo{author}{Ray, S.}, \bibinfo{author}{Sinha,
    S.}, \bibinfo{author}{Shlyapnikov, G.~V.} \& \bibinfo{author}{Santos, L.}
  \newblock \bibinfo{title}{One-dimensional quasicrystals with power-law
    hopping}.
  \newblock \emph{\bibinfo{journal}{Phys. Rev. Lett.}}
  \textbf{\bibinfo{volume}{123}}, \bibinfo{pages}{025301}
  (\bibinfo{year}{2019}).
  \newblock
  \urlprefix\url{https://link.aps.org/doi/10.1103/PhysRevLett.123.025301}.
  \newblock \doi{10.1103/PhysRevLett.123.025301} .

\bibitem{PhysRevB.75.155111}
  \bibinfo{author}{Oganesyan, V.} \& \bibinfo{author}{Huse, D.~A.}
  \newblock \bibinfo{title}{Localization of interacting fermions at high
    temperature}.
  \newblock \emph{\bibinfo{journal}{Phys. Rev. B}} \textbf{\bibinfo{volume}{75}},
  \bibinfo{pages}{155111} (\bibinfo{year}{2007}).
  \newblock \urlprefix\url{https://link.aps.org/doi/10.1103/PhysRevB.75.155111}.
  \newblock \doi{10.1103/PhysRevB.75.155111} .

\bibitem{PhysRevLett.110.084101}
  \bibinfo{author}{Atas, Y.~Y.}, \bibinfo{author}{Bogomolny, E.},
  \bibinfo{author}{Giraud, O.} \& \bibinfo{author}{Roux, G.}
  \newblock \bibinfo{title}{Distribution of the ratio of consecutive level
    spacings in random matrix ensembles}.
  \newblock \emph{\bibinfo{journal}{Phys. Rev. Lett.}}
  \textbf{\bibinfo{volume}{110}}, \bibinfo{pages}{084101}
  (\bibinfo{year}{2013}).
  \newblock
  \urlprefix\url{https://link.aps.org/doi/10.1103/PhysRevLett.110.084101}.
  \newblock \doi{10.1103/PhysRevLett.110.084101} .

\bibitem{PhysRevB.97.125116}
  \bibinfo{author}{Roy, N.} \& \bibinfo{author}{Sharma, A.}
  \newblock \bibinfo{title}{Entanglement contour perspective for ``strong
    area-law violation'' in a disordered long-range hopping model}.
  \newblock \emph{\bibinfo{journal}{Phys. Rev. B}} \textbf{\bibinfo{volume}{97}},
  \bibinfo{pages}{125116} (\bibinfo{year}{2018}).
  \newblock \urlprefix\url{https://link.aps.org/doi/10.1103/PhysRevB.97.125116}.
  \newblock \doi{10.1103/PhysRevB.97.125116} .

\bibitem{PhysRevLett.98.110601}
  \bibinfo{author}{Buonsante, P.} \& \bibinfo{author}{Vezzani, A.}
  \newblock \bibinfo{title}{Ground-state fidelity and bipartite entanglement in
    the bose-hubbard model}.
  \newblock \emph{\bibinfo{journal}{Phys. Rev. Lett.}}
  \textbf{\bibinfo{volume}{98}}, \bibinfo{pages}{110601}
  (\bibinfo{year}{2007}).
  \newblock
  \urlprefix\url{https://link.aps.org/doi/10.1103/PhysRevLett.98.110601}.
  \newblock \doi{10.1103/PhysRevLett.98.110601} .

\bibitem{PhysRevA.77.032111}
  \bibinfo{author}{Chen, S.}, \bibinfo{author}{Wang, L.}, \bibinfo{author}{Hao,
    Y.} \& \bibinfo{author}{Wang, Y.}
  \newblock \bibinfo{title}{Intrinsic relation between ground-state fidelity and
    the characterization of a quantum phase transition}.
  \newblock \emph{\bibinfo{journal}{Phys. Rev. A}} \textbf{\bibinfo{volume}{77}},
  \bibinfo{pages}{032111} (\bibinfo{year}{2008}).
  \newblock \urlprefix\url{https://link.aps.org/doi/10.1103/PhysRevA.77.032111}.
  \newblock \doi{10.1103/PhysRevA.77.032111} .

\bibitem{PhysRevB.76.180403}
  \bibinfo{author}{Yang, M.-F.}
  \newblock \bibinfo{title}{Ground-state fidelity in one-dimensional gapless
    models}.
  \newblock \emph{\bibinfo{journal}{Phys. Rev. B}} \textbf{\bibinfo{volume}{76}},
  \bibinfo{pages}{180403} (\bibinfo{year}{2007}).
  \newblock \urlprefix\url{https://link.aps.org/doi/10.1103/PhysRevB.76.180403}.
  \newblock \doi{10.1103/PhysRevB.76.180403} .

\bibitem{PhysRevE.98.062137}
  \bibinfo{author}{Rossini, D.} \& \bibinfo{author}{Vicari, E.}
  \newblock \bibinfo{title}{Ground-state fidelity at first-order quantum
    transitions}.
  \newblock \emph{\bibinfo{journal}{Phys. Rev. E}} \textbf{\bibinfo{volume}{98}},
  \bibinfo{pages}{062137} (\bibinfo{year}{2018}).
  \newblock \urlprefix\url{https://link.aps.org/doi/10.1103/PhysRevE.98.062137}.
  \newblock \doi{10.1103/PhysRevE.98.062137} .

\bibitem{PhysRevB.80.014403}
  \bibinfo{author}{Zhao, J.-H.} \& \bibinfo{author}{Zhou, H.-Q.}
  \newblock \bibinfo{title}{Singularities in ground-state fidelity and quantum
    phase transitions for the kitaev model}.
  \newblock \emph{\bibinfo{journal}{Phys. Rev. B}} \textbf{\bibinfo{volume}{80}},
  \bibinfo{pages}{014403} (\bibinfo{year}{2009}).
  \newblock \urlprefix\url{https://link.aps.org/doi/10.1103/PhysRevB.80.014403}.
  \newblock \doi{10.1103/PhysRevB.80.014403} .

\bibitem{PhysRevA.89.033625}
  \bibinfo{author}{\L\k{a}cki, M.}, \bibinfo{author}{Damski, B.} \&
  \bibinfo{author}{Zakrzewski, J.}
  \newblock \bibinfo{title}{Numerical studies of ground-state fidelity of the
    bose-hubbard model}.
  \newblock \emph{\bibinfo{journal}{Phys. Rev. A}} \textbf{\bibinfo{volume}{89}},
  \bibinfo{pages}{033625} (\bibinfo{year}{2014}).
  \newblock \urlprefix\url{https://link.aps.org/doi/10.1103/PhysRevA.89.033625}.
  \newblock \doi{10.1103/PhysRevA.89.033625} .

\bibitem{PhysRevLett.18.1049}
  \bibinfo{author}{Anderson, P.~W.}
  \newblock \bibinfo{title}{Infrared catastrophe in fermi gases with local
    scattering potentials}.
  \newblock \emph{\bibinfo{journal}{Phys. Rev. Lett.}}
  \textbf{\bibinfo{volume}{18}}, \bibinfo{pages}{1049--1051}
  (\bibinfo{year}{1967}).
  \newblock \urlprefix\url{https://link.aps.org/doi/10.1103/PhysRevLett.18.1049}.
  \newblock \doi{10.1103/PhysRevLett.18.1049} .

\bibitem{PhysRevB.92.054203}
  \bibinfo{author}{Vasseur, R.} \& \bibinfo{author}{Moore, J.~E.}
  \newblock \bibinfo{title}{Multifractal orthogonality catastrophe in
    one-dimensional random quantum critical points}.
  \newblock \emph{\bibinfo{journal}{Phys. Rev. B}} \textbf{\bibinfo{volume}{92}},
  \bibinfo{pages}{054203} (\bibinfo{year}{2015}).
  \newblock \urlprefix\url{https://link.aps.org/doi/10.1103/PhysRevB.92.054203}.
  \newblock \doi{10.1103/PhysRevB.92.054203} .

\bibitem{PhysRevB.92.220201}
  \bibinfo{author}{Deng, D.-L.}, \bibinfo{author}{Pixley, J.~H.},
  \bibinfo{author}{Li, X.} \& \bibinfo{author}{Das~Sarma, S.}
  \newblock \bibinfo{title}{Exponential orthogonality catastrophe in
    single-particle and many-body localized systems}.
  \newblock \emph{\bibinfo{journal}{Phys. Rev. B}} \textbf{\bibinfo{volume}{92}},
  \bibinfo{pages}{220201} (\bibinfo{year}{2015}).
  \newblock \urlprefix\url{https://link.aps.org/doi/10.1103/PhysRevB.92.220201}.
  \newblock \doi{10.1103/PhysRevB.92.220201} .

\bibitem{Cosco_2018}
  \bibinfo{author}{Cosco, F.} \emph{et~al.}
  \newblock \bibinfo{title}{Statistics of orthogonality catastrophe events in
    localised disordered lattices}.
  \newblock \emph{\bibinfo{journal}{New Journal of Physics}}
  \textbf{\bibinfo{volume}{20}}~(7), \bibinfo{pages}{073041}
  (\bibinfo{year}{2018}).
  \newblock \urlprefix\url{https://doi.org/10.1088/1367-2630/aad10b}.
  \newblock \doi{10.1088/1367-2630/aad10b} .

\bibitem{PhysRevLett.122.040604}
  \bibinfo{author}{Tonielli, F.}, \bibinfo{author}{Fazio, R.},
  \bibinfo{author}{Diehl, S.} \& \bibinfo{author}{Marino, J.}
  \newblock \bibinfo{title}{Orthogonality catastrophe in dissipative quantum
    many-body systems}.
  \newblock \emph{\bibinfo{journal}{Phys. Rev. Lett.}}
  \textbf{\bibinfo{volume}{122}}, \bibinfo{pages}{040604}
  (\bibinfo{year}{2019}).
  \newblock
  \urlprefix\url{https://link.aps.org/doi/10.1103/PhysRevLett.122.040604}.
  \newblock \doi{10.1103/PhysRevLett.122.040604} .

\bibitem{PhysRevLett.93.266402}
  \bibinfo{author}{Levine, G.~C.}
  \newblock \bibinfo{title}{Entanglement entropy in a boundary impurity model}.
  \newblock \emph{\bibinfo{journal}{Phys. Rev. Lett.}}
  \textbf{\bibinfo{volume}{93}}, \bibinfo{pages}{266402}
  (\bibinfo{year}{2004}).
  \newblock
  \urlprefix\url{https://link.aps.org/doi/10.1103/PhysRevLett.93.266402}.
  \newblock \doi{10.1103/PhysRevLett.93.266402} .

\bibitem{Peschel_2005}
  \bibinfo{author}{Peschel, I.}
  \newblock \bibinfo{title}{Entanglement entropy with interface defects}.
  \newblock \emph{\bibinfo{journal}{Journal of Physics A: Mathematical and
      General}} \textbf{\bibinfo{volume}{38}}~(20), \bibinfo{pages}{4327--4335}
  (\bibinfo{year}{2005}).
  \newblock \urlprefix\url{https://doi.org/10.1088/0305-4470/38/20/002}.
  \newblock \doi{10.1088/0305-4470/38/20/002} .

\bibitem{Edwards_1972}
  \bibinfo{author}{Edwards, J.~T.} \& \bibinfo{author}{Thouless, D.~J.}
  \newblock \bibinfo{title}{Numerical studies of localization in disordered
    systems}.
  \newblock \emph{\bibinfo{journal}{Journal of Physics C: Solid State Physics}}
  \textbf{\bibinfo{volume}{5}}~(8), \bibinfo{pages}{807--820}
  (\bibinfo{year}{1972}).
  \newblock \urlprefix\url{https://doi.org/10.1088%2F0022-3719%2F5%2F8%2F007}.
    \newblock \doi{10.1088/0022-3719/5/8/007} .

  \bibitem{PhysRevB.96.045123}
    \bibinfo{author}{Mohammad, P.} \& \bibinfo{author}{Montakhab, A.}
    \newblock \bibinfo{title}{Sensitivity of the entanglement spectrum to boundary
      conditions as a characterization of the phase transition from delocalization
      to localization}.
    \newblock \emph{\bibinfo{journal}{Phys. Rev. B}} \textbf{\bibinfo{volume}{96}},
    \bibinfo{pages}{045123} (\bibinfo{year}{2017}).
    \newblock \urlprefix\url{https://link.aps.org/doi/10.1103/PhysRevB.96.045123}.
    \newblock \doi{10.1103/PhysRevB.96.045123} .

  \bibitem{PhysRevB.103.035136}
    \bibinfo{author}{Pouranvari, M.} \& \bibinfo{author}{Liou, S.-F.}
    \newblock \bibinfo{title}{Characterizing many-body localization via state
      sensitivity to boundary conditions}.
    \newblock \emph{\bibinfo{journal}{Phys. Rev. B}}
    \textbf{\bibinfo{volume}{103}}, \bibinfo{pages}{035136}
    (\bibinfo{year}{2021}).
    \newblock \urlprefix\url{https://link.aps.org/doi/10.1103/PhysRevB.103.035136}.
    \newblock \doi{10.1103/PhysRevB.103.035136} .

  \bibitem{PhysRevLett.101.256802}
    \bibinfo{author}{Hashimoto, K.} \emph{et~al.}
    \newblock \bibinfo{title}{Quantum hall transition in real space: From localized
      to extended states}.
    \newblock \emph{\bibinfo{journal}{Phys. Rev. Lett.}}
    \textbf{\bibinfo{volume}{101}}, \bibinfo{pages}{256802}
    (\bibinfo{year}{2008}).
    \newblock
    \urlprefix\url{https://link.aps.org/doi/10.1103/PhysRevLett.101.256802}.
    \newblock \doi{10.1103/PhysRevLett.101.256802} .

  \bibitem{PhysRevLett.65.88}
    \bibinfo{author}{Dunlap, D.~H.}, \bibinfo{author}{Wu, H.-L.} \&
    \bibinfo{author}{Phillips, P.~W.}
    \newblock \bibinfo{title}{Absence of localization in a random-dimer model}.
    \newblock \emph{\bibinfo{journal}{Phys. Rev. Lett.}}
    \textbf{\bibinfo{volume}{65}}, \bibinfo{pages}{88--91}
    (\bibinfo{year}{1990}).
    \newblock \urlprefix\url{https://link.aps.org/doi/10.1103/PhysRevLett.65.88}.
    \newblock \doi{10.1103/PhysRevLett.65.88} .

  \bibitem{Bovier_1992}
    \bibinfo{author}{Bovier, A.}
    \newblock \bibinfo{title}{Perturbation theory for the random dimer model}.
    \newblock \emph{\bibinfo{journal}{Journal of Physics A: Mathematical and
        General}} \textbf{\bibinfo{volume}{25}}~(5), \bibinfo{pages}{1021--1029}
    (\bibinfo{year}{1992}).
    \newblock \urlprefix\url{https://doi.org/10.1088/0305-4470/25/5/011}.
    \newblock \doi{10.1088/0305-4470/25/5/011} .

  \bibitem{PhysRevB.69.085109}
    \bibinfo{author}{Sedrakyan, T.}
    \newblock \bibinfo{title}{Localization-delocalization transition in a presence
      of correlated disorder: The random dimer model}.
    \newblock \emph{\bibinfo{journal}{Phys. Rev. B}} \textbf{\bibinfo{volume}{69}},
    \bibinfo{pages}{085109} (\bibinfo{year}{2004}).
    \newblock \urlprefix\url{https://link.aps.org/doi/10.1103/PhysRevB.69.085109}.
    \newblock \doi{10.1103/PhysRevB.69.085109} .

  \bibitem{PhysRevB.48.16347}
    \bibinfo{author}{Datta, P.~K.}, \bibinfo{author}{Giri, D.} \&
    \bibinfo{author}{Kundu, K.}
    \newblock \bibinfo{title}{Nature of states in a random-dimer model:
      Bandwidth-scaling analysis}.
    \newblock \emph{\bibinfo{journal}{Phys. Rev. B}} \textbf{\bibinfo{volume}{48}},
    \bibinfo{pages}{16347--16356} (\bibinfo{year}{1993}).
    \newblock \urlprefix\url{https://link.aps.org/doi/10.1103/PhysRevB.48.16347}.
    \newblock \doi{10.1103/PhysRevB.48.16347} .

  \bibitem{PhysRevB.56.1170}
    \bibinfo{author}{Farchioni, R.} \& \bibinfo{author}{Grosso, G.}
    \newblock \bibinfo{title}{Electronic transport for random dimer-trimer model
      hamiltonians}.
    \newblock \emph{\bibinfo{journal}{Phys. Rev. B}} \textbf{\bibinfo{volume}{56}},
    \bibinfo{pages}{1170--1174} (\bibinfo{year}{1997}).
    \newblock \urlprefix\url{https://link.aps.org/doi/10.1103/PhysRevB.56.1170}.
    \newblock \doi{10.1103/PhysRevB.56.1170} .

  \bibitem{PhysRevB.100.195109}
    \bibinfo{author}{Pouranvari, M.} \& \bibinfo{author}{Abouie, J.}
    \newblock \bibinfo{title}{Entanglement conductance as a characterization of a
      delocalized-localized phase transition in free fermion models}.
    \newblock \emph{\bibinfo{journal}{Phys. Rev. B}}
    \textbf{\bibinfo{volume}{100}}, \bibinfo{pages}{195109}
    (\bibinfo{year}{2019}).
    \newblock \urlprefix\url{https://link.aps.org/doi/10.1103/PhysRevB.100.195109}.
    \newblock \doi{10.1103/PhysRevB.100.195109} .

  \bibitem{PhysRevLett.114.146601}
    \bibinfo{author}{Ganeshan, S.}, \bibinfo{author}{Pixley, J.~H.} \&
    \bibinfo{author}{Das~Sarma, S.}
    \newblock \bibinfo{title}{Nearest neighbor tight binding models with an exact
      mobility edge in one dimension}.
    \newblock \emph{\bibinfo{journal}{Phys. Rev. Lett.}}
    \textbf{\bibinfo{volume}{114}}, \bibinfo{pages}{146601}
    (\bibinfo{year}{2015}).
    \newblock
    \urlprefix\url{https://link.aps.org/doi/10.1103/PhysRevLett.114.146601}.
    \newblock \doi{10.1103/PhysRevLett.114.146601} .

  \bibitem{Dom_nguez_Castro_2019}
    \bibinfo{author}{Dom{\'{\i}}nguez-Castro, G.~A.} \& \bibinfo{author}{Paredes,
      R.}
    \newblock \bibinfo{title}{The aubry{\textendash}andr{\'{e}} model as a
      hobbyhorse for understanding the localization phenomenon}.
    \newblock \emph{\bibinfo{journal}{European Journal of Physics}}
    \textbf{\bibinfo{volume}{40}}~(4), \bibinfo{pages}{045403}
    (\bibinfo{year}{2019}).
    \newblock \urlprefix\url{https://doi.org/10.1088/1361-6404/ab1670}.
    \newblock \doi{10.1088/1361-6404/ab1670} .

  \bibitem{PhysRevB.101.174203}
    \bibinfo{author}{Cookmeyer, T.}, \bibinfo{author}{Motruk, J.} \&
    \bibinfo{author}{Moore, J.~E.}
    \newblock \bibinfo{title}{Critical properties of the ground-state
      localization-delocalization transition in the many-particle aubry-andr\'e
      model}.
    \newblock \emph{\bibinfo{journal}{Phys. Rev. B}}
    \textbf{\bibinfo{volume}{101}}, \bibinfo{pages}{174203}
    (\bibinfo{year}{2020}).
    \newblock \urlprefix\url{https://link.aps.org/doi/10.1103/PhysRevB.101.174203}.
    \newblock \doi{10.1103/PhysRevB.101.174203} .

  \bibitem{PhysRevB.101.024202}
    \bibinfo{author}{Riddell, J.} \& \bibinfo{author}{S\o{}rensen, E.~S.}
    \newblock \bibinfo{title}{Out-of-time-order correlations in the quasiperiodic
      aubry-andr\'e model}.
    \newblock \emph{\bibinfo{journal}{Phys. Rev. B}}
    \textbf{\bibinfo{volume}{101}}, \bibinfo{pages}{024202}
    (\bibinfo{year}{2020}).
    \newblock \urlprefix\url{https://link.aps.org/doi/10.1103/PhysRevB.101.024202}.
    \newblock \doi{10.1103/PhysRevB.101.024202} .

  \bibitem{laug}
    \bibinfo{author}{Anderson, E.} \emph{et~al.}
    \newblock \emph{\bibinfo{title}{{LAPACK} Users' Guide}}
    \bibinfo{edition}{Third} edn (\bibinfo{publisher}{Society for Industrial and
      Applied Mathematics}, \bibinfo{address}{Philadelphia, PA},
    \bibinfo{year}{1999}).

  \bibitem{fraxanet2022localization}
    \bibinfo{author}{Fraxanet, J.}, \bibinfo{author}{Bhattacharya, U.},
    \bibinfo{author}{Grass, T.}, \bibinfo{author}{Lewenstein, M.} \&
    \bibinfo{author}{Dauphin, A.}
    \newblock \bibinfo{title}{Localization and multifractal properties of the
      long-range kitaev chain in the presence of an aubry-andr\'e-harper
      modulation} (\bibinfo{year}{2022}).
    \newblock \eprint{2201.05458}.

  \bibitem{PhysRevB.83.075105}
    \bibinfo{author}{Biddle, J.}, \bibinfo{author}{Priour, D.~J.},
    \bibinfo{author}{Wang, B.} \& \bibinfo{author}{Das~Sarma, S.}
    \newblock \bibinfo{title}{Localization in one-dimensional lattices with
      non-nearest-neighbor hopping: Generalized anderson and aubry-andr\'e models}.
    \newblock \emph{\bibinfo{journal}{Phys. Rev. B}} \textbf{\bibinfo{volume}{83}},
    \bibinfo{pages}{075105} (\bibinfo{year}{2011}).
    \newblock \urlprefix\url{https://link.aps.org/doi/10.1103/PhysRevB.83.075105}.
    \newblock \doi{10.1103/PhysRevB.83.075105} .

  \end{thebibliography}

\end{document}